# Active Learning for Out-of-class Activities by Using Interactive Mobile Apps

Muztaba Fuad
Dept. of Computer Science
Winston-Salem State University
Winston-Salem, NC, USA
fuadmo@wssu.edu

Monika Akbar
Dept. of Computer Science
University of Texas-El Paso
El-Paso, TX, USA
makbar@utep.edu

Lynn Zubov
Dept. of Education
Winston-Salem State University
Winston-Salem, NC, USA
zubovl@wssu.edu

*Abstract*—Keeping students engaged with the course content outside the classroom is a challenging task. Since learning during undergraduate years occurs not only as student engagement in class, but also during out-of-class activities, we need to redesign and reinvent such activities for this and future generation of students. Although active learning has been used widely to improve in-class student learning and engagement, its usage outside the classroom is not widespread and researched. Active learning is often not utilized for out-of-class activities and traditional unsupervised activities are used mostly to keep students engaged in the content after they leave the classroom. Although there has been tremendous research performed to improve student learning and engagement in the classroom, there are a few pieces of researches on improving out-of-class learning and student engagement. This poster will present an approach to redesign the traditional out-of-class activities with the help of mobile apps, which are interactive and adaptive, and will provide personalization to satisfy student's needs outside the classroom so that optimal learning experience can be achieved.

## I. Significance and Relevance

Today's college freshman students are not as prepared for college work as their predecessors [1]. This problem is exacerbated when they spend fewer hours studying outside the classroom [1]. A 2003 study performed by University of California-LA's Graduate School of Education and Information Studies [2] shows that this trend starts well before the students show up in college. Such trends provide some insight into why students are under prepared and are less engaged in study. Another trend that started showing up [3]-[6] in the last decade or so is the gradual increase of the amount of hours student spend on their jobs. This is mainly because of rising tuition and living expenses and of other socio-cultural reasons [4]-[5]. Students spending less time studying outside the class and working more for money are the two new problems that most institutions are currently facing. Therefore, it is crucial to investigate approaches that will ensure effective usage of the time that the students have for studying outside the classroom. Regardless of the factors leading to students working more, the impact of studying less outside the classroom has been studied extensively in the recent years [5]-[6] and since the objectives of those studies are complex, a generalized conclusion on the relationship between hours employed and academic performance cannot be made. Therefore, instead of focusing on this issue, this research will focus on how students can best utilize the time that they allocate to perform out-of-class activities and how we can provide such activities in more active, adaptive and structured way that facilitates greater engagement and improved student learning.

Although most college guidelines and literature [7]-[8] states that students should study at home around 2-3 hours per credit hour of class per week, students rarely follow that [3]-[4]. Our own experience also supports those findings. From the summative course evaluation survey data (Fig. 1) in one of the target course for this research, student's self-reported hours for enrolled credit, study time outside class and employment shows that they are not spending enough time studying outside the class as suggested.

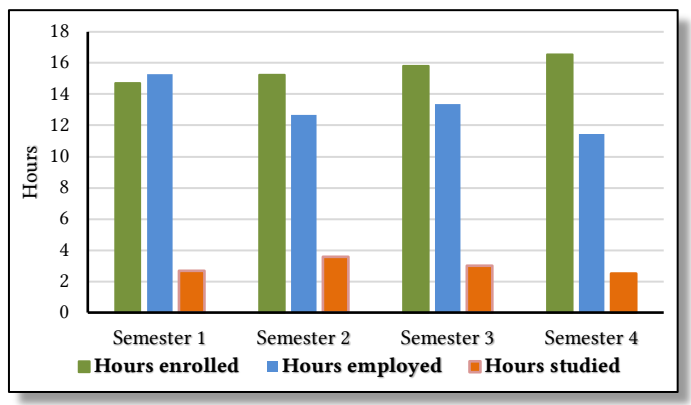

Fig. 1. Student self-reported times for a target course.

This poster is going to introduce a mobile-based system, being developed, called dysgu (means 'learning' in Welsh), which will provide students with interactive and motivating out-of-class activities. Dysgu provides personalization and adaptation to satisfy student's specific needs to optimize the learning experience. Dysgu employs a novel approach to conduct out-of-class active learning activities enhanced with five unique factors. First, the activities are interactive [9] and smaller than traditional out-of-class activities allowing for instructional scaffolding and can be deployed more frequently and with higher expectations to promote self-fulfilling prophecy. Second, dysgu utilizes mobile technology-based learning environment to deliver and administer such activities, as current generation of students are active users of this technology. Third, it supports adaptation and personalization to address student's lifestyle needs. Fourth, social networking is incorporated in the learning environment, allowing students to anonymously participate in interaction and to view their own progress compared to their peers. Finally, gamification components are used to encourage students to participate and become more engaged in the activities. The goal of dysgu is to make the time that student spend on out-of-class activities more productive, so that they are more engaged (by interactive nature of dysgu), more focused (by adaptive and personalized nature of dysgu) and motivated (by social and gamified nature of dysgu); which in turn may lead to improved learning experiences.

In recent years, the concept of flipped classes [10] have become very popular for its perceived benefits in student learning and engagement. However, there are concerns [11] related to this new pedagogy, especially that it might not work for every type of students, especially those who come from socially disadvantaged background. Dysgu can address those concerns by its nature of pervasive access, interactivity, adaptability and motivation-oriented environment to augment flipped classes to make it more productive for a diverse class of students.

## II. CURRENT STATUS

Dysgu is currently under development and planned to be deployed in classes in Fall of 2018. The faculty-side (Fig. 2(a)) of the software is developed using Java and the student-side (Fig 2(b)) is developed using Android (the iOS version is planned). One freshman and one sophomore level course are identified where dysgu will be deployed. Both of these courses assess mathematics and problem-solving skills, which will allow us to utilize different cognitivist and constructivist principles to enhance student's skills. Since generally students at this level struggles with factors related to transitioning from high school to college, the intended intervention will have better impact on their learning and engagement as older students are already matured enough to handle such factors.

## III. CONCLUSION

With the help of Dysgu, the proposed model will exploit students' usage of technology and mobile devices for interaction and learning outside the classroom. Dysgu will facilitate active and blended learning, where participation is part compulsory-part interest driven, is learner-centric and where learning is being evaluated continuously. By having a guided learning environment and by using mobile technology, we believe that students can be steered more effectively once they leave the classroom and expect students to maintain more focus on the course content and ultimately to learn and retain information better.

This poster will present experience and insight gained in this research and showcase a synopsis of the three-year project. The poster will present the overall system design and software architecture of the Dysgu system, descriptions of key system elements, description and examples of interactive problems and course modules, evaluation strategies and current status of the research.

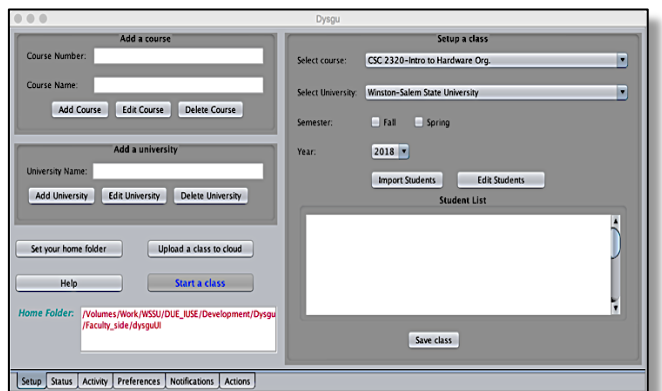

(a)

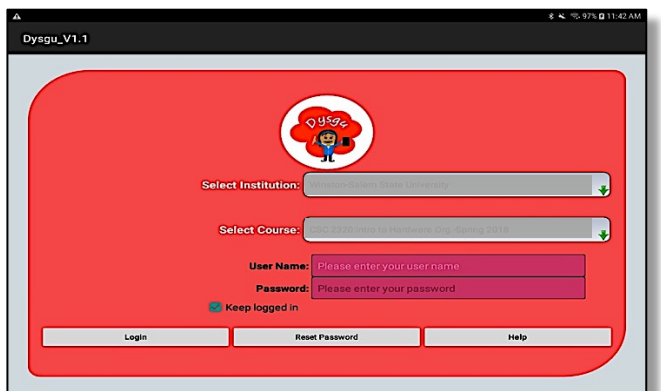

(b)

Fig. 2. Sample of Dysgu user interfaces.

## ACKNOWLEDGMENTS

This research is supported by National Science Foundation grant # 1712030 and 1712073.

## REFERENCES

[1] Philip S. Babcock and Mindy Marks (2010), The Falling Time Cost of College: Evidence from Half a Century of Time Use Data, NBER Working Paper No. 15954.
[2] Higher Education Research Institute. (2003). The official press release for the American freshmen 2002. UCLA Press.
[3] Gose, B. (1998). More freshmen than ever appear disengaged from their studies, survey finds. The Chronicle of Higher Education, A37–A39.
[4] Nonis, S.A. and Hudson, G.I. (2006). Academic Performance of College Students: Influence of Time Spent Studying and Working. Journal of Education for Business, 81(3), 151-159.
[5] Miller K., Danner, F., and Staten, R. (2008). Relationship of Work Hours with Selected Health Behaviors and Academic Progress Among a College Student Cohort. Journal of American College Health, 56(6), 675-679.
[6] Callender, C. (2008). The impact of term-time employment on higher education students' academic attainment and achievement. Journal of Education Policy, 23(4), 359-377.
[7] Bennett, J. (2000) Hints on How to Succeed in College Classes, http://www.jeffreybennett.com/pdf/How_to_Succeed_general.pdf.
[8] Academic Success Center, How Many Hours Do I Need to Study? USU, https://www.usu.edu/asc/assistance/pdf/estimate_study_hours.pdf.
[9] Anonymous reference
[10] Bishop, J.L. & Verleger, M.A. (2013) "The Flipped Classroom: A Survey of the Research," 120th American Society of Engineering Education Annual Conference & Exposition, Atlanta, Georgia, USA, June 23-26.
[11] Weimer M. (2014), A Few Concerns about the Rush to Flip, Faculty Focus, Magna Publication.